\documentclass[12pt]{article}

\usepackage{scicite}
\usepackage{graphicx}

\usepackage{times}

\newenvironment{sciabstract}{%
\begin{quote} \bf}
{\end{quote}}

\newcounter{lastnote}

\title{Detection of a branched alkyl molecule in the interstellar 
medium: \textit{iso}-propyl cyanide}

\author
{Arnaud Belloche,$^{1\ast}$ Robin T. Garrod,$^{2}$ Holger S.~P. 
M{\"u}ller,$^{3}$ Karl M. Menten$^{1}$\\
\\
\normalsize{$^{1}$Max-Planck-Institut f{\"u}r Radioastronomie, Auf dem H{\"u}gel 69, 53121 Bonn, Germany}\\
\normalsize{$^{2}$Center for Radiophysics and Space Research, Cornell University, Ithaca, NY 14853-6801, USA}\\
\normalsize{$^{3}$I. Physikalisches Institut, Universit{\"a}t zu K{\"o}ln, Z{\"u}lpicher Str. 77, 50937 K{\"o}ln, Germany}\\
\\
\normalsize{$^\ast$To whom correspondence should be addressed; E-mail:  belloche@mpifr-bonn.mpg.de.}
}

\date{}

\begin{document} 


\baselineskip24pt


\maketitle


\begin{sciabstract}
The largest non-cyclic molecules detected in the interstellar medium (ISM) 
are organic with a straight-chain carbon backbone. We report an 
interstellar detection of a branched alkyl molecule, \textit{iso}-propyl 
cyanide (\textit{i}-C$_3$H$_7$CN), with an abundance 0.4 times that of its 
straight-chain structural isomer. This 
detection suggests that branched carbon-chain molecules may be generally
abundant in the ISM. Our astrochemical model indicates that both isomers are 
produced within or upon dust grain ice mantles through the 
addition of molecular 
radicals, albeit via differing reaction pathways. 
The production of \textit{iso}-propyl cyanide appears to require the
addition of a functional group to a non-terminal carbon in the chain. Its
detection therefore bodes well for the presence in the ISM of amino acids, for 
which such side-chain structure is a key characteristic.
\end{sciabstract}



Around 180 molecules have been detected so far in 
the interstellar medium (ISM)\cite{molspace}. Apart from the fullerenes that
are cyclic, the largest of these interstellar molecules are organic with a 
straight-chain 
carbon backbone. However, more complex molecules have been 
identified in meteorites found on Earth, including more than 80 amino 
acids\cite{Elsila07} -- the building blocks of proteins. The composition of 
these meteoritic amino acids suggests that they or their direct 
precursors have an interstellar origin\cite{Botta02}. Chemistry at work in the
ISM may thus be capable of producing organic molecules more complex than those 
detected so far and thus of great importance to astrobiology. However, the 
degree of complexity that may be reached in the ISM is still an open question, 
as well as how widespread these complex organic molecules are in our 
galaxy. 

Our previous observations of the star-forming region 
Sagittarius B2(N) -- 
hereafter Sgr~B2(N) -- yielded the first interstellar detection of the 
straight-chain organic molecule \textit{normal}-propyl cyanide 
(\textit{n}-C$_3$H$_7$CN), the largest molecule yet detected in this 
source\cite{Belloche09}. This spectral line survey, conducted using the 
30-m single-dish radio telescope of the Institut de 
Radioastronomie Millim\'etrique (IRAM), provided continuous spectral 
coverage throughout the 3-mm transmission window of Earth's 
atmosphere\cite{Belloche13}. It allowed the identification of other 
complex organic species such as ethyl formate\cite{Belloche09} and 
aminoacetonitrile\cite{Belloche08}, a possible precursor of the amino acid 
glycine. On the basis of the success of this single-dish line survey, we 
performed a survey in the same waveband by using the Atacama Large 
Millimeter/submillimeter 
Array (ALMA), resulting in an increase of more than one order of magnitude in 
both sensitivity and angular resolution. This interferometric project, called 
EMoCA (Exploring Molecular Complexity with ALMA), aims to 
decipher the 
molecular content of Sgr~B2(N) in order to test the predictions of 
astrochemical numerical simulations and to gain insight into the chemical 
processes at work in the ISM.

Sgr~B2 is the most massive star-forming region in our 
galaxy. It is located close to the Galactic Center, which is
$8.34 \pm 0.16$~kpc from the Sun\cite{Reid14}. Sgr~B2 contains two 
main sites of 
star formation, Sgr~B2(N) and Sgr~B2(M) that, since the early 1970s, have 
turned out to be the best hunting ground for complex organic molecules in the 
ISM. Their immense hydrogen column densities that signify 
large quantities of gas enable the detection of low-abundance 
species. Sgr~B2(N) 
itself contains two dense, compact, hot cores that are separated by about 
$5''$ ($\sim$ 40,000~AU in projection)\cite{Belloche08}.

Propyl cyanide (C$_3$H$_7$CN, hereafter PrCN) is the smallest alkyl cyanide 
that exists in several distinct isomers (Fig.~1)\textbf{:} the chain 
isomer \textit{normal}- or \textit{n}-PrCN (also known as butyronitrile or 
1-cyanopropane) and the branched isomer \textit{iso}- or 
\textit{i}-PrCN 
(also known as \textit{iso}-butyronitrile or 2-cyanopropane). 
\textit{n}-PrCN is the smallest alkyl cyanide that exists in several 
distinct conformations. The CN group can be attached to the terminal C of the 
propyl group in the CCC plane, trans to the CCC chain, leading to the 
\textit{anti} conformer, also known as \textit{trans} (Fig.~1C); it can 
also be attached to the propyl group rotated by $\pm 120^\circ$ with respect to 
the CCC plane, leading to the \textit{gauche} conformer (Fig.~1B). 
The rotational 
spectrum of \textit{i}-PrCN, previously only studied to a limited extent in 
the microwave region, has recently been recorded extensively in the laboratory
from the microwave to the submillimeter wave region along with a 
redetermination of the dipole moment\cite{i-PrCN_rot_2011}. 

We used ALMA in 2012 to perform a full spectral line survey toward 
Sgr~B2(N) in the 3~mm atmospheric window between 84 and 
111~GHz\cite{SupplMat}. We identified the detected lines by 
modeling the molecular 
emission under the assumption of local thermodynamic equilibrium.
By employing predictions from the Cologne Database for Molecular 
Spectroscopy\cite{CDMS2}, we assigned emission features
to \textit{i}-PrCN or \textit{n}-PrCN\cite{SupplMat}.
To interpret the astronomical detections, we performed numerical
simulations of the chemistry occuring during the evolution of a hot
core\cite{SupplMat}.

Many spectral lines are detected in the ALMA data toward both of the 
hot cores 
embedded in Sgr~B2(N). These spectra are very close to the confusion limit,
i.e., signal from a spectral line is detected in nearly every 
spectral channel.
The lines are narrower toward the northern, less prominent hot core 
(full width at half maximum, FWHM,
$\sim 5$~km~s$^{-1}$) than toward the southern, more prominent one. The 
detection of faint lines from rare species is therefore easier toward the 
former and we focus on this one in the present work. We constructed a 
preliminary model of the emission of all molecules previously detected, on the
basis of our analysis of the previous single-dish survey of 
Sgr~B2(N)\cite{Belloche13}. In this way, the risk of mis-assigning a line to a 
new species is reduced. About 50 and 120 transitions of \textit{i}-PrCN and 
\textit{n}-PrCN, respectively, are detected toward the 
northern hot core (Fig.~2A and B and Figs. S1 and S2). On the basis 
of this model, we selected the least contaminated transitions 
and produced contour maps of their intensity integrated over their 
line profile (Fig.~2C and E and Figs. S3 and 
S4). From these maps,  we derive a deconvolved angular size of 
$1.0'' \pm 0.3''$ 
(FWHM) for the region where both species emit. We used the population diagram 
method\cite{Goldsmith99} to estimate a rotation temperature 
of $153 \pm 12$~K (SEM), which characterizes
the emission of both molecules\cite{SupplMat}.

With the size and temperature derived above and a linewidth 
measurement of 5 km~s$^{-1}$, we obtain a good fit to all 
transitions of \textit{i}-PrCN and \textit{n}-PrCN detected toward the northern
hot core of Sgr~B2(N). After correction for the contribution of vibrationally 
excited states\cite{SupplMat}, we derive column densities of 
$7.2 \pm 1.4 \times 10^{16}$~cm$^{-2}$ and $1.8 \pm 0.4 \times 10^{17}$~cm$^{-2}$
(SEM), respectively,
which yields an abundance ratio [\textit{i}-PrCN]/[\textit{n}-PrCN] of 
$0.40 \pm 0.06$ (SEM). The latter uncertainty assumes the same 
source size and rotation temperature for both isomers. 
With the H$_2$ column density derived from the continuum 
emission (\textit{\citen{SupplMat}}, Fig.~2D), we deduce 
average abundances relative to H$_2$ of $1.3 \pm 0.2 \times 10^{-8}$ for 
\textit{i}-PrCN and $3.2 \pm 0.5 \times 10^{-8}$ (SEM) for \textit{n}-PrCN.
The latter uncertainties assume the same rotation and dust temperatures, and
take neither possible contamination of the continuum emission by free-free 
emission nor uncertainties on the dust properties into account.

The recently developed chemical kinetics model MAGICKAL
(Model for Astrophysical Gas and Ice Chemical Kinetics And Layering,
\textit{\citen{Garrod13}}) was 
used to simulate the time-dependent chemistry of the source. The model
begins with a cold collapse phase, during which abundant ice 
mantles, composed of simple H, O, C, and N-bearing molecules, are formed on 
dust-grain surfaces. 
The cold stage is followed by a warm-up stage, during which the dust and gas 
temperature rises from $\sim 8$~K to 400~K. The majority of complex organic 
molecule formation occurs during this stage, through the addition of simple 
and complex radicals within and upon the ice mantles. 

The model produces time-dependent chemical abundances (with respect to
H$_2$) for various cyanide molecules (Fig. 3). The model temperatures at which
each molecule's peak abundance is attained (Table S2) may be considered
representative of the excitation temperature at which most of the emission from
each molecule would occur, assuming local thermodynamic equilibrium.
The desorption temperature of each 
PrCN isomer is $\sim 150$~K, with peak gas-phase abundances achieved at 160~K. 
This agrees well with the rotation temperatures that we determine
from observational data for these molecules.

The majority of each of the two PrCN isomers form in or on the 
dust-grain ices at around 55--75~K in the model. However, the model 
also indicates that, 
whereas many similar chemical pathways are open to both \textit{i}-PrCN and 
\textit{n}-PrCN, the dominant formation route in each case is different. The 
greatest contribution to \textit{i}-PrCN production comes from the reaction of 
CN radicals (which are accreted from the gas) with the 
CH$_3$CHCH$_3$ radical. The 
latter derives from the earlier gas-phase formation of C$_3$, which is 
hydrogenated and stored on the grains as C$_3$H$_2$, C$_3$H$_4$, and 
C$_3$H$_6$. The addition of atomic hydrogen to propylene (C$_3$H$_6$) 
significantly favors the production of CH$_3$CHCH$_3$, whose radical site 
lies at the secondary carbon atom\cite{Curran06}.

Because the production -- either by hydrogen addition or abstraction 
processes -- of radicals such as CH$_2$CH$_2$CH$_3$ and CH$_2$CH$_2$CN (whose 
radical site is at the primary carbon atom) is strongly disfavored versus 
their equivalent \textit{iso-} forms, we find that the dominant 
formation mechanism for 
\textit{n}-PrCN is the addition of C$_2$H$_5$ and CH$_2$CN -- 
a process which has no equivalent for the production of \textit{i}-PrCN. The 
radicals form through the abstraction of hydrogen by OH, from C$_2$H$_6$ 
and CH$_3$CN, respectively. \textit{i}-PrCN production dominates all reaction 
mechanisms for which parallel processes are available to both isomers.

Although the overall peak gas-phase values of \textit{i}-PrCN and 
\textit{n}-PrCN produced by the models are similar, they show a slight bias 
toward \textit{i}-PrCN production (2.2:1), rather than the observed bias toward 
\textit{n}-PrCN (0.4:1). This may be caused by the poorly-defined rates for 
barrier-mediated surface reactions, such as 
\hbox{H + C$_3$H$_6$ $\rightarrow$ CH$_3$CHCH$_3$} and 
\hbox{OH + CH$_3$CN $\rightarrow$ CH$_2$CN + H$_2$O}, for which only gas-phase 
rates have been measured, and whose behavior may be somewhat different on an 
ice surface. 

Amid the growing understanding that complex organic molecules could form on 
the surface of dust grains, the formation of branched alkyl molecules in the 
ISM was suggested theoretically in the 1980s\cite{Millar88} but 
no such molecules were detected until now. The detection of a 
branched alkyl molecule in Sgr~B2, with an abundance similar to its 
straight-chain isomer, indicates a further divergence between the 
chemistry of star-forming regions like Sgr~B2 and quiescent 
regions. These less active regions
seem to produce only linear molecules -- the largest one known to date being 
HC$_{11}$N\cite{Bell97}. 
The detection of a branched alkyl molecule also suggests a 
further link between interstellar chemistry and the molecular composition of 
meteorites for which branched amino acids are even found to dominate over their 
straight-chain isomers\cite{Cronin83}. 
The inherent bias toward the production of secondary rather than primary 
radical sites on precursor radicals suggests that branched molecules may be 
prevalent, and indeed dominant, in star forming regions where chemistry of 
sufficient complexity is reached. The detection of the next member of the 
alkyl cyanide series, \textit{n}-butyl cyanide (\textit{n}-C$_4$H$_9$CN), and 
its three branched isomers would allow the testing of this conjecture.





\noindent\textbf{Acknowledgments:} We thank D. Petry and E. Humphreys at 
the European ALMA Regional Center (ARC), F. Gueth at the IRAM ARC node, and D. 
Muders at the MPIfR for their help with the reduction of the ALMA data; S. 
Bardeau, S. Maret, J. Pety, M. Lonjaret, and P. Hily-Blant for 
the new 
developments implemented in the Weeds software; and M. Koerber for help 
with preparing Fig.~1. This work made use of the NIST 
Chemical Kinetics Database.
This work has been supported in part by the Deutsche Forschungsgemeinschaft
through the collaborative research grant SFB~956 ''Conditions and Impact 
of Star Formation'', project area B3. 
H.S.P.M. is very grateful to the Bundesministerium f\"ur Bildung und Forschung
for support through projects FKZ 50OF0901 (ICC HIFI \textit{Herschel}) 
and 05A11PK3 (ALMA ARC Node).
R.T.G. acknowledges support from the NASA Astrophysics Theory Program through 
grant NNX11AC38G.
This paper makes use of the following ALMA data: ADS/JAO.ALMA\#2011.0.00017.S. 
ALMA is a partnership of ESO (representing its member states), NSF (USA), and 
NINS (Japan), together with NRC (Canada) and NSC and ASIAA (Taiwan), in 
cooperation with the Republic of Chile. The Joint ALMA Observatory is operated 
by ESO, AUI/NRAO, and NAOJ.
The interferometric data are available in the ALMA archive at
https://almascience.eso.org/aq/.
The chemical model input files are located at 
www.astro.cornell.edu/$\sim$rgarrod/resources.


\vspace{2ex}
\noindent\textbf{Supplementary Materials}\\
www.sciencemag.org/content/345/6204/1584/suppl/DC1\\
Materials and Methods\\
Figs. S1 to S6\\
Tables S1 and S2\\
References (\textit{\citen{n-PrCN_rot_1982}}--\textit{\citen{Hu97}})


\clearpage

\begin{figure}
\centerline{\resizebox{0.35\hsize}{!}{\includegraphics[angle=0]{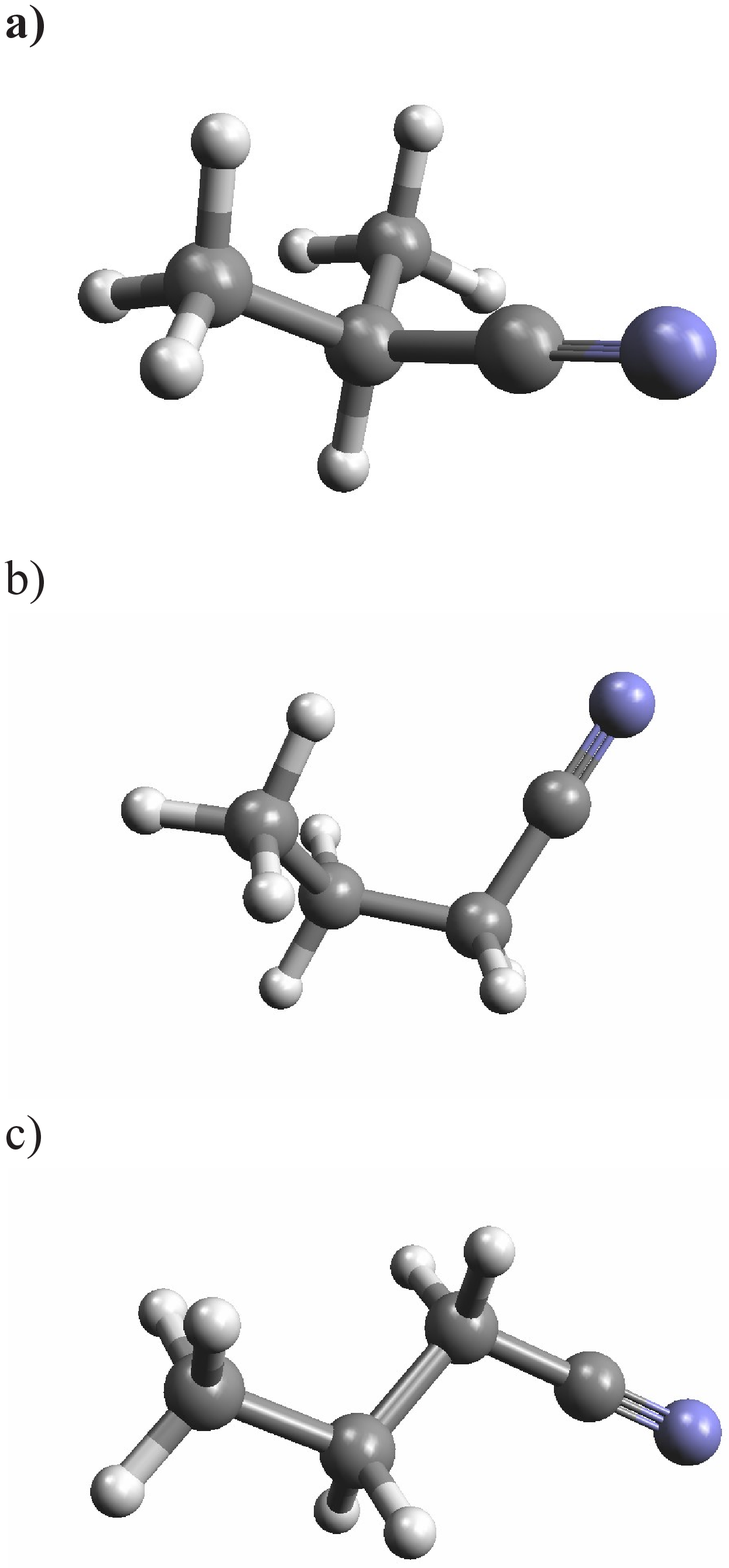}}}
\caption{Models of propyl cyanide isomers and conformers.
(\textbf{A}) \textit{iso}-propyl cyanide (\textit{i}-C$_3$H$_7$CN).
(\textbf{B}) \textit{gauche}-\textit{normal}-propyl cyanide 
(\textit{g-n}-C$_3$H$_7$CN).
(\textbf{C}) \textit{anti}-\textit{normal}-propyl cyanide  
(\textit{a-n}-C$_3$H$_7$CN). 
C, H, and N atoms are indicated by gray, small light gray, and blue spheres, 
respectively.}
\label{f:molecules}
\end{figure}

\begin{figure}
\centerline{\resizebox{0.78\hsize}{!}{\includegraphics[angle=270]{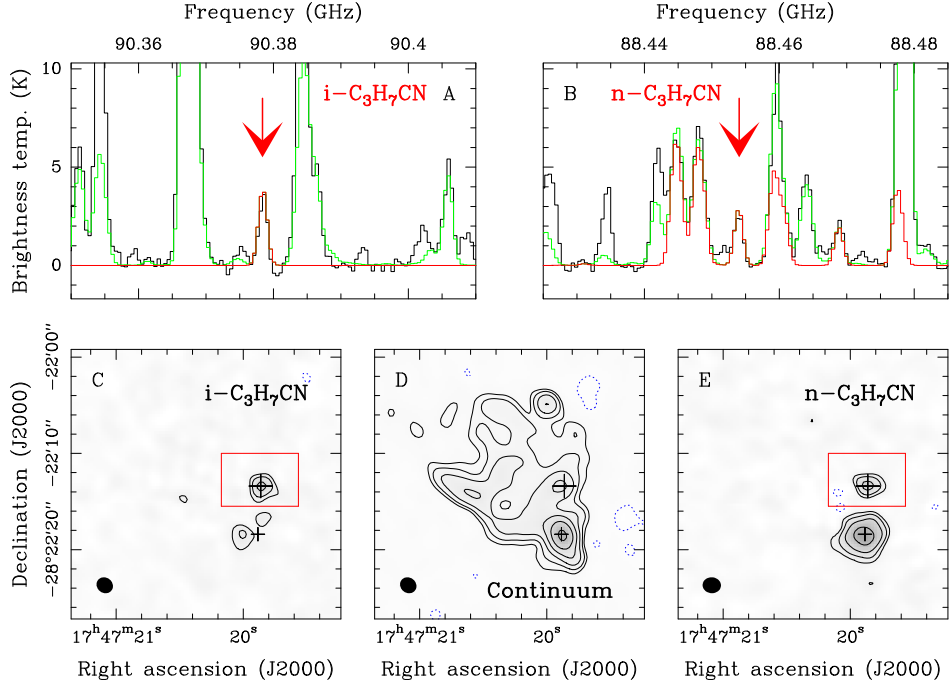}}}
\caption{Examples of transitions of \textit{i}-PrCN and \textit{n}-PrCN 
toward the northern hot core of Sgr~B2(N). 
(\textbf{A} and \textbf{B}) Continuum-subtracted spectrum observed with ALMA in
black, the preliminary model including all identified
molecules in green, and the synthetic spectra of \textit{i}-PrCN 
and \textit{n}-PrCN, respectively, in red.
(\textbf{C} and \textbf{E}) Integrated intensity maps of the transitions of 
\textit{i}-PrCN and \textit{n}-PrCN marked with a red arrow in 
(A) and (B), respectively. The continuum emission at 
90.5~GHz is shown in (\textbf{D}). The negative contour (dotted blue) is 
$-3\sigma$ and the positive contours (black) are $2^i \times 3\sigma$, with 
$i$ an integer starting at 0 and $\sigma$ the root-mean-square noise level 
(23~mJy~beam$^{-1}$~km~s$^{-1}$, 
8.3~mJy~beam$^{-1}$, and 
20~mJy~beam$^{-1}$~km~s$^{-1}$ 
with half-power beam widths of 1.8$''$$\times$1.6$''$, 1.8$''$$\times$1.6$''$,
and 1.9$''$$\times$1.6$''$ in (C), (D), 
and (E), respectively). The large cross indicates the position of the 
northern hot core as traced by both molecules. The smaller cross marks the 
position of the main hot core that has a lower systemic velocity, which 
implies that the contours outside the red box do not trace the emission of 
\textit{i}-PrCN and \textit{n}-PrCN in (C) and (E), 
respectively. The black ellipses show the size of the respective synthetic 
beams.}
\label{f:specmap}
\end{figure}

\begin{figure}
\centerline{\includegraphics[angle=0]{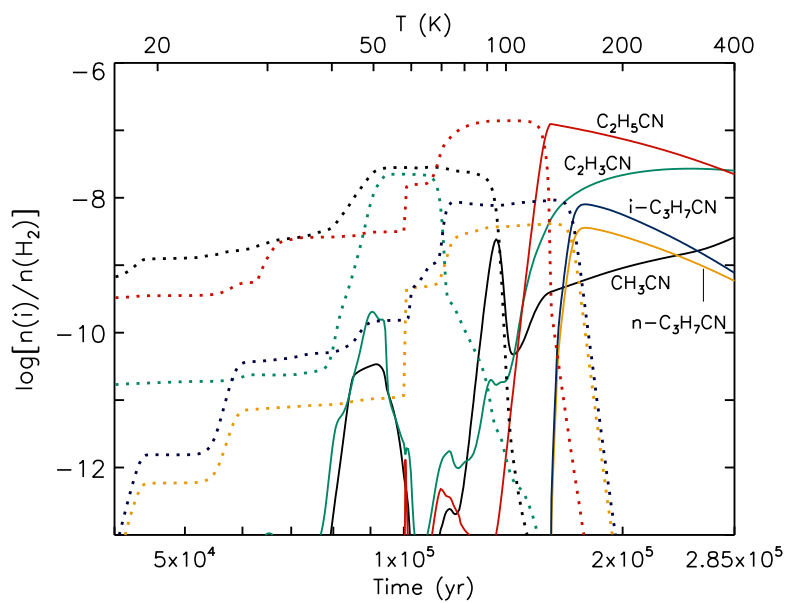}}
\caption{Simulated fractional chemical abundances of alkyl cyanides 
with respect to molecular hydrogen, H$_2$. These abundances represent
the warm-up phase of 
hot-core evolution. Solid lines indicate gas-phase species; dotted lines of 
the same colour indicate the same species in the solid phase. The main phase 
change from solid to gas for each molecule is caused by thermal desorption 
from the grain surfaces, according to species-specific binding energies.}
\label{f:model}
\end{figure}

\clearpage

\centerline{\textbf{\LARGE Supplementary Materials}}

\section*{Materials and Methods}

\setcounter{figure}{0}
\setcounter{table}{0}
\renewcommand{\thefigure}{S\arabic{figure}}
\renewcommand{\thetable}{S\arabic{table}}

\paragraph*{Observations.} The observations were 
carried out with the Atacama Large Millimeter/submillimeter Array (ALMA) 
between August and October 2012. The full frequency range between 84.1 and 
110.7 GHz was covered with four slightly overlapping spectral setups. Each 
setup was observed between one and four times for a total of ten observing 
tracks lasting between $\sim 0.6$ and 2~h each, but only the longest track of
each setup was used in this work. The number of antennas for the selected 
tracks varied between 22 and 26. The size (HPBW) of the primary beam varies 
between $74''$ at 84~GHz and $56''$ at 111~GHz. The field was centered at 
$\alpha_{\rm J2000}$=17$^{\rm h}$47$^{\rm m}$19.87$^{\rm s}$, 
$\delta_{\rm J2000}$=$-28^\circ$22$'$16$''$,
half way between the two hot cores embedded in Sgr~B2(N) that are separated by
$4.9''$ in the north-south direction. The channel spacing was 244 kHz, but 
the spectra were smoothed to 488 kHz (1.7--1.3 km~s$^{-1}$). Each setup 
consists of four spectral windows of 1875 MHz each, two per sideband. The 
separation between the centers of the lower and upper sidebands is 12 GHz. 

\paragraph*{Data reduction.} The data were calibrated and imaged with CASA 
(Version 4.2.0, r28322). The bandpass calibration was carried out on the 
quasar B1730-130. The absolute flux density scale was derived from Neptune 
(except for the selected track of the second setup that used Titan). The phase 
and amplitude calibrations were performed on the quasar J1700-261. The track 
of the third setup was affected by an astrometry problem with Sgr~B2(N) being 
shifted by $(\Delta\alpha,\Delta\delta) \sim (0.5'',0.2'')$ with respect to 
the other setups. This shift may be due to an inacurrate calibration of the 
atmospheric phase fluctuations due to the low elevation of the phase 
calibrator during the observations of this particular track. The offset was 
approximately compensated for by 
modifying the visibilities of the phase calibrator with the CASA task 
\textit{fixvis} before the phase calibration. After this correction,
the relative positional accuracy of the four tracks used in this work is
on the order of $\pm 0.1''$ in both right ascension and declination.
Three or four iterations of self-calibration were performed using a strong 
spectral line detected toward Sgr~B2(N) in each setup.
The images were produced with the CLEAN deconvolution algorithm in CASA.
The size (HPBW) of the synthesized beam depends on the setup and the sideband 
but is typically $2.0'' \times 1.4''$ (minimum $1.6'' \times 1.2''$, maximum 
$3.0'' \times 1.4''$). The median rms noise level reached for 
each selected track varies between 4.4 and 9.9 mJy/beam. 
Based on the redundancies of the tracks and the spectral overlap between the 
setups, we estimate the absolute calibration uncertainty on the flux density 
to be on the order of 15\%. 

The spectra toward and around the two hot cores embedded in Sgr~B2(N) are
full of lines and close to the confusion limit. It is thus difficult to 
separate the line emission from the continuum emission in a systematic way for
the full data cubes, but it is a necessary step to produce separate line and 
continuum maps. For each spectral window of each setup, we selected six groups 
of few channels that seemed to be free of strong line emission. 
A first-order baseline was fitted to these selected channels and the result of
the fit was used to split each data cube into two cubes, one for the line 
emission and one for the continuum emission. Given the difference in systemic 
velocity between the two hot cores ($\sim 10$~km~s$^{-1}$, see
\textit{\citen{Belloche13}}), we selected different sets of channels for the 
northern and southern parts of the field.

\paragraph*{Laboratory spectroscopy.}
Predictions of the rotational spectra of \textit{i}-PrCN and \textit{n}-PrCN 
were taken from the CDMS\cite{CDMS2}, version 1 and 2, 
respectively. The predictions of \textit{i}-PrCN were based on spectroscopic 
parameters derived from microwave to submillimeter wave 
measurements\cite{i-PrCN_rot_2011}. The dipole moment components were also 
redetermined in that work. Belloche et al.\cite{Belloche09} derived 
spectroscopic parameters of both \textit{gauche} and \textit{anti} conformers 
of \textit{n}-PrCN by combining data from two microwave 
studies\cite{n-PrCN_rot_1982,n-PrCN_rot1_1988} with millimeter and 
submillimeter data from Wlodarczak et al.\cite{n-PrCN_rot2_1988}.

Wlodarczak et al.\cite{n-PrCN_rot2_1988} also determined the \textit{a}- and 
\textit{b}-dipole moment components for both conformers of \textit{n}-PrCN and 
determined the \textit{anti} conformer to be lower in energy than the 
\textit{gauche} 
conformer by $1.1 \pm 0.3$~kJ/mol, in contrast to an electron diffraction 
study\cite{n-PrCN_e-Diff_2000} and to a more recent and much more accurate 
determination by low-temperature infrared measurements that determined the 
\textit{gauche} conformer to be lower in energy by 
$0.48 \pm 0.04$~kJ/mol\cite{n-PrCN_vib_2001}. The results of our ALMA 
observations are fully consistent with the latter energy difference.
The reverse energy ordering in 
the millimeter wave study\cite{n-PrCN_rot2_1988} may, at least in part, be 
due to an underestimation of the strong \textit{a}-dipole moment component of 
the \textit{anti} conformer. In fact, quantum-chemical calculations on 
\textit{i}-PrCN, \textit{a-n}-PrCN and 
\textit{cyclo}-PrCN\cite{i-PrCN_rot_2011} suggest that the \textit{a}-dipole 
moment component of \textit{a-n}-PrCN is around 4.0~D instead of the measured 
$3.597 \pm 0.059$~D\cite{n-PrCN_rot2_1988}. This underestimation leads to an 
overestimation of the amount of \textit{a-n}-PrCN by about one fourth because 
the intensity of a 
transition scales with the square of the dipole moment component. The 
experimental value of the weaker \textit{b}-dipole moment component of almost 
1.0~D may be correct, but could also be slightly underestimated. Similar 
quantum-chemical calculations on the dipole moment components of 
\textit{g-n}-PrCN by one of us (HSPM) provide good 
agreement with the experiment for the \textit{a}- and \textit{b}-components 
and suggest that neglecting the small \textit{c}-component of $\sim$0.45~D 
leads to only small errors. Nevertheless, this component was used in the 
second version of the \textit{n}-PrCN CDMS entry.

A considerable part of such complex molecules resides in excited vibrational 
states at the elevated temperatures determined for the propyl cyanides. In 
particular, low-lying vibrational states are of importance. We use data 
from Durig et al.\cite{i-PrCN_etc_FIR_1972} for 
\textit{i}-PrCN and from Durig et al.\cite{n-PrCN_vib_2001} for 
\textit{n}-PrCN. Contributions from vibrationally excited states increase the 
partition function 
values at 150~K by factors of 1.64, 2.09, and 2.36 for \textit{i}-PrCN, 
\textit{g-n}-PrCN, and \textit{a-n}-PrCN, respectively.

\paragraph*{Line identification and spectral modeling.}
The line identification in the ALMA spectra was performed following the 
same strategy as for our previous, single-dish, spectral-line survey of 
Sgr~B2(N) with the IRAM 30~m telescope\cite{Belloche13}. The main difference 
is that we used Weeds\cite{Maret11} (version of April 2014) which is part of 
the CLASS software (see http://www.iram.fr/IRAMFR/GILDAS). We built our own 
Weeds database containing mostly JPL\cite{Pickett98} and CDMS\cite{CDMS2} 
entries, plus a number of private entries as in Belloche et 
al.\cite{Belloche13}. For this work, we focused on the northern hot core 
embedded in Sgr~B2(N) (position P2 in ref.~\textit{\citen{Belloche08}}). The 
emission of each molecule across the full 
3~mm atmospheric window is modeled assuming local thermodynamic equilibrium
(LTE, which means that the excitation temperature of the transitions is equal
to the rotational temperature of the molecule and the kinetic temperature of 
the gas) with five parameters: column density, temperature, source size, 
velocity offset, and linewidth. We model each spectral window of each observed 
setup separately to account for the varying angular resolution.

A preliminary model fitting the emission detected with ALMA was constructed 
based on the parameters derived for all molecules identified in our 
single-dish survey of Sgr~B2(N). This model has not been fully optimized for 
ALMA yet but it is already sufficiently reliable to allow for detecting new 
molecules in the ALMA spectra. Figures~S1 and S2 show the transitions 
of \textit{i}-PrCN and \textit{n}-PrCN, respectively, identified in the ALMA 
spectrum of the northern hot core. Only the transitions that are not severely 
blended with transitions of other species are shown in these figures. 
Integrated intensity maps were computed for the transitions that are well 
fitted by our LTE model in these figures, i.e. the transitions that do not 
suffer from blends with identified or unidentified species. These maps are 
shown in Figs.~S3 and S4. We stress the fact that the emission 
was integrated over only a few channels (typically $\sim 11$~km~s$^{-1}$) 
covering each transition detected toward the northern hot core 
(FWHM $\sim 5$~km~s$^{-1}$). This means that, in these figures, the contours 
around the main hot core, which has a systemic velocity lower by 
$\sim 10$~km~s$^{-1}$ compared to the northern hot core\cite{Belloche13}, 
mainly trace other species than \textit{i}-PrCN or \textit{n}-PrCN. The emission
toward the northern hot core was fitted with an elliptical Gaussian for each 
map. The results are similar for both molecules.
Ignoring the transitions from the third setup that has the largest beam 
and for which the absolute astrometry was readjusted, we derive equatorial 
offsets $(\Delta\alpha, \Delta\delta)_{\rm J2000} = 
(-0.1'' \pm 0.1''$, $2.6'' \pm 0.1''$) for the peak position with 
respect to the phase center and a size of $1.0'' \pm 0.3''$ (FWHM).

Finally, we constructed population diagrams\cite{Goldsmith99} for 
\textit{i}-PrCN and \textit{n}-PrCN using the transitions that are not too 
severely blended with emission from other identified species. Figures~S5A and 
S6A show the population diagrams derived using the observed 
integrated intensities of \textit{i}-PrCN and \textit{n}-PrCN, respectively. 
Figures~S5B and S6B
show the same after subtracting the (model) contribution of all contaminating 
molecules that are included in our full LTE synthetic spectrum, like we did
for our previous single-dish detection of \textit{n}-PrCN\cite{Belloche09}. In 
each panel, the LTE model of \textit{i}-PrCN or \textit{n}-PrCN is 
shown in red. The observed data points that remain well above the synthetic 
data points in Figs.~S5B and S6B are most likely contaminated by 
emission from still unidentified species. Nevertheless, a linear fit to these
data points can give a rough estimate of the rotation temperature 
characterizing the emission of the molecules, provided the transitions are 
optically thin (which is verified to be true a posteriori). In such a diagram, 
the slope of the linear fit is inversely proportional to the rotation
temperature. We derive $T_{\rm ex} = 154 \pm 99$~K for \textit{i}-PrCN and 
$153 \pm 12$~K (SEM) for \textit{n}-PrCN. The larger uncertainty for 
the former is due to the narrower energy range of its detected transitions. 
We estimate the statistical error on the derived column densities to
be on the order of 20\% (SEM), with 10\% coming from the uncertainty on the 
(\textit{n}-PrCN) rotation temperature. This uncertainty does not take into 
account the uncertainty on the source size and the overall calibration 
uncertainty that should however both affect the column densities of 
\textit{i}-PrCN and \textit{n}-PrCN in the same way.

\paragraph*{Continuum emission.}
The peak flux density measured toward the northern hot core at 98.8~GHz is
0.24~Jy/1.8$''$$\times$1.3$''$-beam. Assuming a dust mass opacity of 
0.01~cm$^{2}$~g$^{-1}$ at 230~GHz\cite{Ossenkopf94}, a dust emissivity index 
$\beta = 1.0$, a mean molecular weight per H$_2$ molecule of 2.8,
a temperature of 150~K, and that all the continuum flux is due to thermal dust
emission, we derive a peak H$_2$ column density, $N_{{\rm H}_2}$, of 
$4.2 \times 10^{24}$~cm$^{-2}$.
For a distance of 8.34~kpc, the measured peak flux density corresponds to a 
mass, $M$, of 415~$M_\odot$ for a region of size 12900~AU (FWHM). Assuming 
spherical symmetry, this translates into a mean free-particle density, $n$, of 
$5.5 \times 10^7$~cm$^{-3}$, or a mean hydrogen density $n_{\rm H}$ of 
$9.3 \times 10^7$~cm$^{-3}$.
We extrapolate these results for the more compact region where the PrCN
emission comes from (FWHM $\sim 1.0''$) by assuming spherical symmetry and a 
density profile proportional to $r^{-1.5}$\cite{Osorio99}. We obtain 
$N_{{\rm H}_2}(1'') = 5.2 \times 10^{24}$~cm$^{-2}$,
$M(1'') = 215$~$M_\odot$,
$n(1'') = 1.1 \times 10^8$~cm$^{-3}$, and 
$n_{\rm H}(1'') = 1.8 \times 10^8$~cm$^{-3}$.
Doing the same for all spectral windows of all setups, we derive the following
average values and standard deviations (rms):
$N_{{\rm H}_2}(1'') = 5.6 \pm 0.4 \times 10^{24}$~cm$^{-2}$,
$M(1'') = 232 \pm 17$~$M_\odot$,
$n(1'') = 1.06 \pm 0.08 \times 10^8$~cm$^{-3}$, and 
$n_{\rm H}(1'') = 1.9 \pm 0.1 \times 10^8$~cm$^{-3}$. The uncertainties are
only statistical and do neither take into account the uncertainties due to the 
assumptions about the geometry and $\beta$, nor possible contamination by
free-free emission.

\paragraph*{Astrochemical numerical simulations.}
The chemical model solves the fully-coupled time-dependent chemical kinetics 
of nearly 700 unique chemical species, and traces their populations in the gas 
phase, on dust-grain/ice surfaces, and within the bulk ice mantles, 
during the collapse and warm-up of the hot core, as described by 
Garrod\cite{Garrod13}.
The chemical network consists of a full gas-phase 
chemistry that includes ion-molecule and neutral-neutral reactions, as well as 
solid-phase atomic- and radical-addition reactions and photodissociation 
processes. The gas and grain chemical phases are coupled through gas-phase
accretion and both thermal and non-thermal desorption mechanisms. The chemical 
kinetics are solved using the modified rate-equation approach\cite{Garrod08},
with a standard ODE solver.

The behavior of the chemical system is determined by the physical conditions 
during the two modeled stages of physical evolution of the hot core. The first 
is a free-fall collapse from a total hydrogen density, $n_{\rm H}$, of 3000 to 
$2 \times 10^8$~cm$^{-3}$. The final density is consistent with the density 
derived for the region in the northern hot core of Sgr~B2(N) 
from which the PrCN emission originates (see above).
Gas temperatures are fixed at 10~K, while dust temperatures fall 
from $\sim$18 to 8~K, according to the visual extinction, $A_V$, as described 
by Garrod \& Pauly\cite{Garrod11}. Dust-grain ice mantles form during this 
stage, through hydrogenation of atoms and simple molecules derived from the 
gas phase. The second stage consists of a physically-static warm up of both 
the gas and the dust, to 400 K, assuming the intermediate warm-up period of 
Garrod\cite{Garrod13}. The majority of complex organic chemistry occurring on 
the dust grains takes place during the warm-up stage, during which the 
mobility of reactive radicals is increased. As temperatures rise, surface 
molecules thermally desorb from the grain/ice surface into the gas phase, 
wherein the most complex molecules are typically destroyed by ion-molecule 
reactions.

The major elemental abundances at the beginning of the collapse stage with 
respect to total hydrogen are 0.09 for He, $1.4 \times 10^{-4}$ for C, 
$7.5 \times 10^{-5}$ for N, and $3.2 \times 10^{-4}$ for O. Ice mantle molecular 
abundances at the end of the collapse phase, with respect to water, are 
56\% for CO, 18\% for CO$_2$, 7.0\% for CH$_3$OH, 2.0\% for CH$_4$, and
18\% for NH$_3$.

During the warm-up stage, reactive radicals on the dust grains can form 
through photodissociation of stable molecules, by H-atom abstraction by other 
radicals such as OH, or by H-atom addition to unsaturated molecules. Rates are 
determined as described by Garrod\cite{Garrod13}. Belloche et 
al.\cite{Belloche09} incorporated formation and destruction mechanisms for 
\textit{n}-PrCN. The new chemical model introduces a range of mechanisms for 
its structural isomer, \textit{i}-PrCN. This is the first time that side-chain 
chemistry has been included in interstellar chemical networks.

Table~S1 shows the main reactions included in the model that lead
to the formation of \textit{n}-PrCN and \textit{i}-PrCN upon/within the 
dust-grain ices. They are all assumed to proceed without a barrier. Only the 
final step is shown for each mechanism. The production of the larger precursor 
radicals involved in the listed reactions can occur through hydrogen 
abstraction from a saturated molecule, e.g. C$_3$H$_8$ (either by OH or 
NH$_2$). Photolysis may also produce such radicals, although other product 
branches are also included in the model for each photolysis event. A set of 
analogous reactions to some of those shown in Table~S1 is also included, which
consist of the addition of CH$_2$ (rather than CH$_3$) to a radical. PrCN is
then formed by the addition of a hydrogen atom. The CH$_2$ mechanisms are of
only minor influence to the results. CH$_3$ can be formed by dissociation of
molecules such as methanol (CH$_3$OH), while CN also has multiple gas-phase
and dust-grain routes to its formation.
Some reactions between stable molecules and radicals other than OH and 
NH$_2$ are included in the chemical network, but these are relatively 
inefficient,
due to significant activation energies, and at interstellar temperatures
would strongly favor the abstraction of atomic hydrogen, rather than
producing more complex molecular structures.

Measured gas-phase rates indicate that hydrogenation of C$_3$H$_6$ and 
H-atom abstraction from C$_3$H$_8$ preferentially result in an unpaired 
electron at the secondary carbon site\cite{Curran06,Mebel99,Hu97}, due to the 
lower activation energy, which favors \textit{i}-PrCN 
production over \textit{n}-PrCN. Hydrogenation and abstraction processes 
involving C$_2$H$_3$CN and C$_2$H$_5$CN, respectively, are treated 
similarly. Photodissociation rates are also assumed with a bias of 10:1 
in favor of a radical site at the secondary carbon, although the 
importance of photodissociation is small compared to the former 
processes. All other rates and binding energies are assumed to be 
identical between the \textit{i}-PrCN and \textit{n}-PrCN forms.
Binding energies of molecules for which experimental data on appropriate
ice surfaces do not exist, including PrCN, are determined by interpolation
of known energies for similar species or functional groups.

\clearpage

\begin{figure}
\centerline{\resizebox{0.76\hsize}{!}{\includegraphics[angle=0]{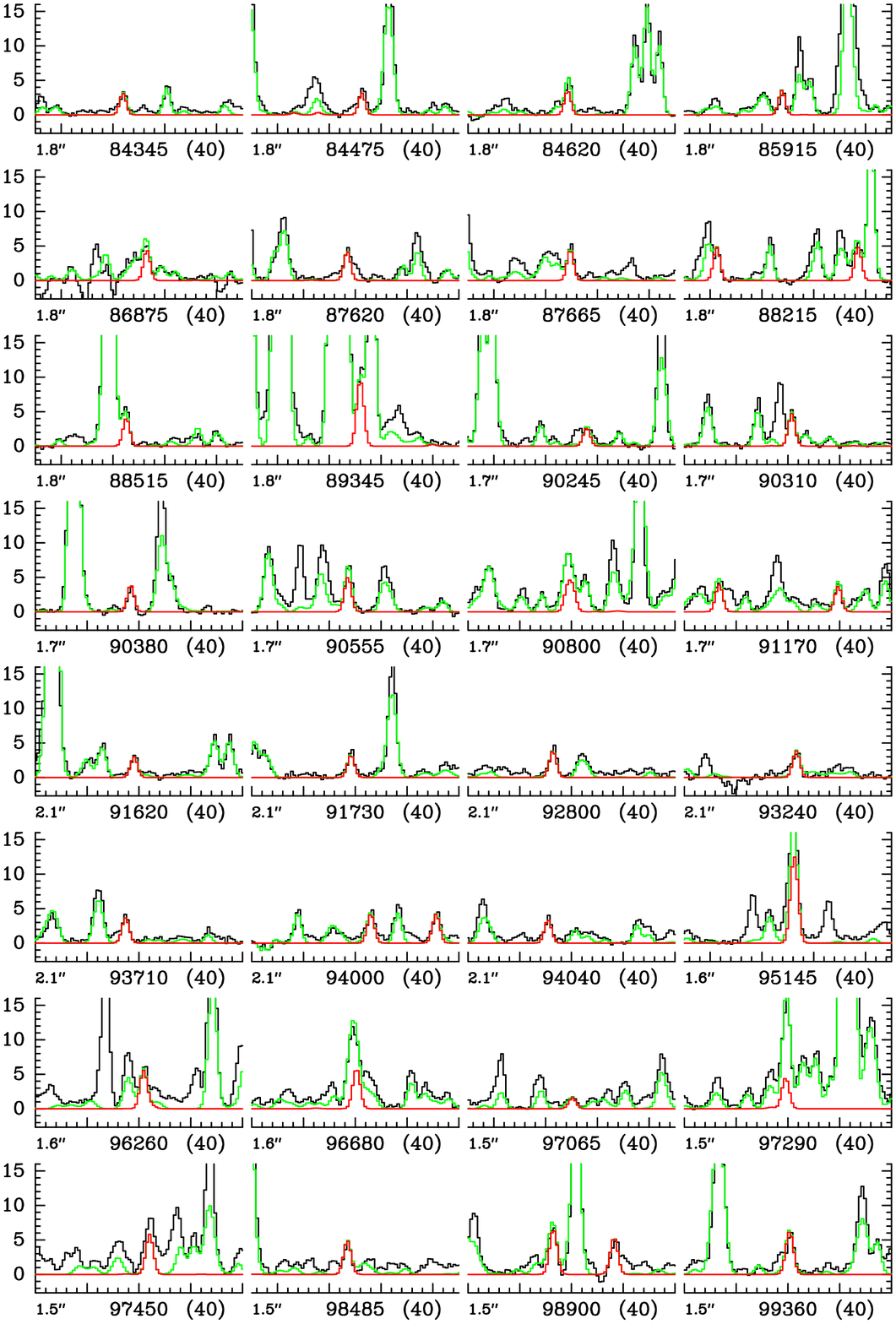}}}
\caption{Transitions of \textit{i}-PrCN detected with ALMA toward 
the northern hot core embedded in Sgr~B2(N). The continuum-subtracted 
observed spectrum is 
shown in black, the synthetic spectrum of \textit{i}-PrCN in red, and the 
preliminary model including all detected molecules in green. The x-axis of 
each panel is labeled with its central frequency and its frequency range in 
parentheses, both in MHz. The angular resolution (mean FWHM of the 
synthesized beam) is also indicated below each panel. The y-axis indicates the 
brightness temperature in K.}
\label{f:spectra_i}
\end{figure}

\begin{figure}
\addtocounter{figure}{-1}
\centerline{\resizebox{0.76\hsize}{!}{\includegraphics[angle=0]{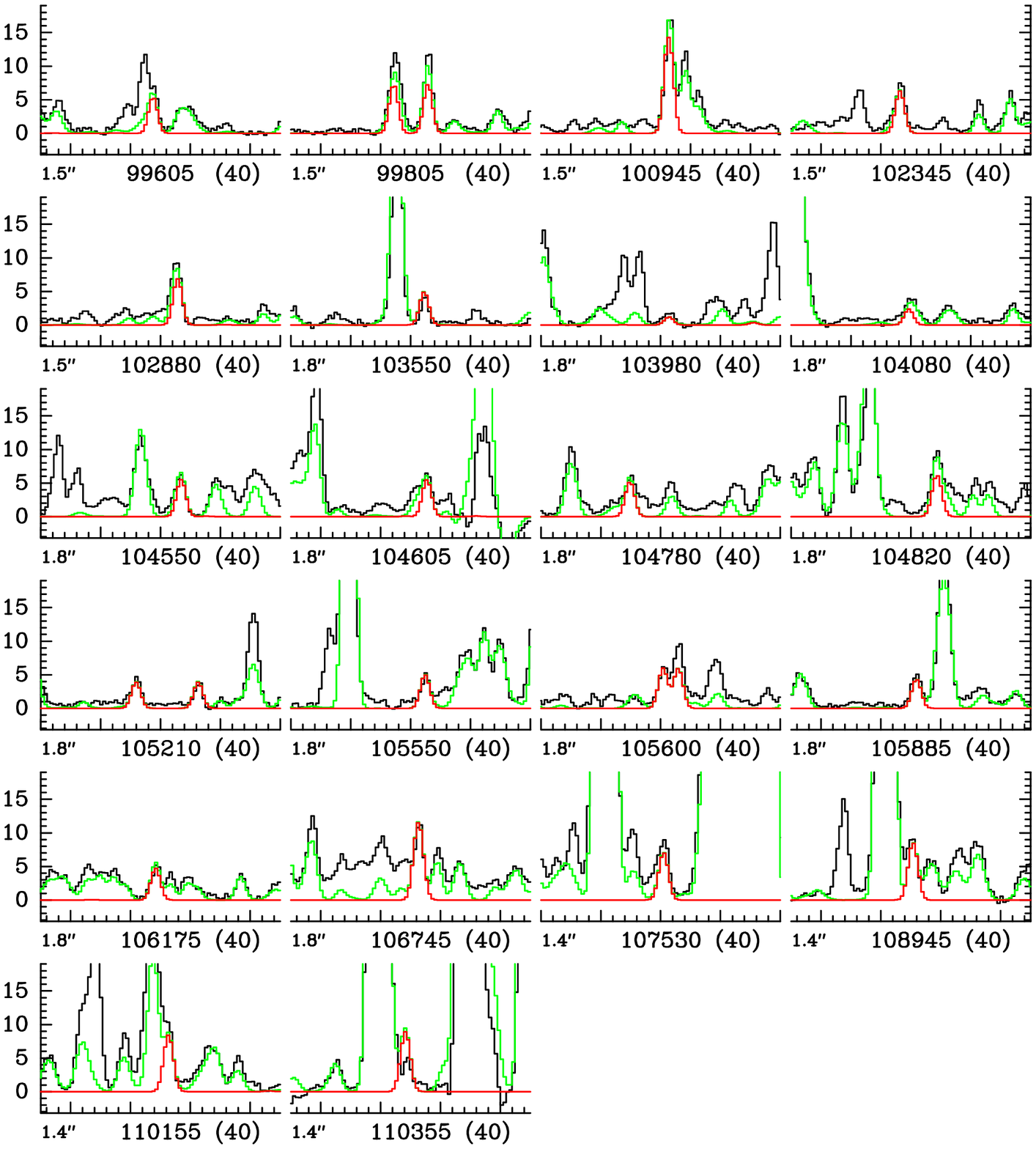}}}
\caption{continued.}
\end{figure}

\clearpage

\begin{figure}
\centerline{\resizebox{0.76\hsize}{!}{\includegraphics[angle=0]{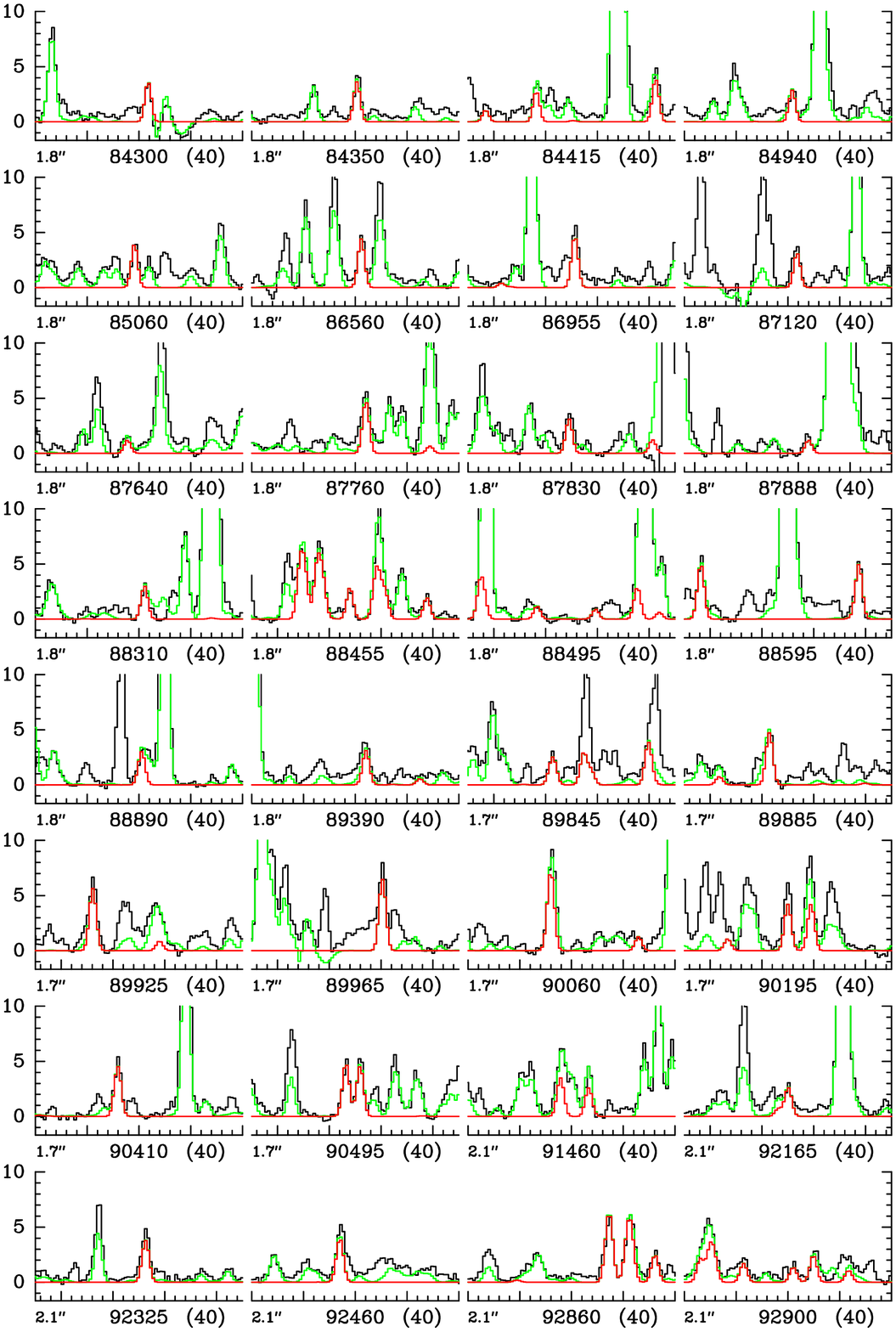}}}
\caption{Transitions of \textit{n}-PrCN detected with ALMA toward 
the northern hot core embedded in Sgr~B2(N). See caption of Fig.~S1 for more 
details.}
\label{f:spectra_n}
\end{figure}

\begin{figure}
\addtocounter{figure}{-1}
\centerline{\resizebox{0.76\hsize}{!}{\includegraphics[angle=0]{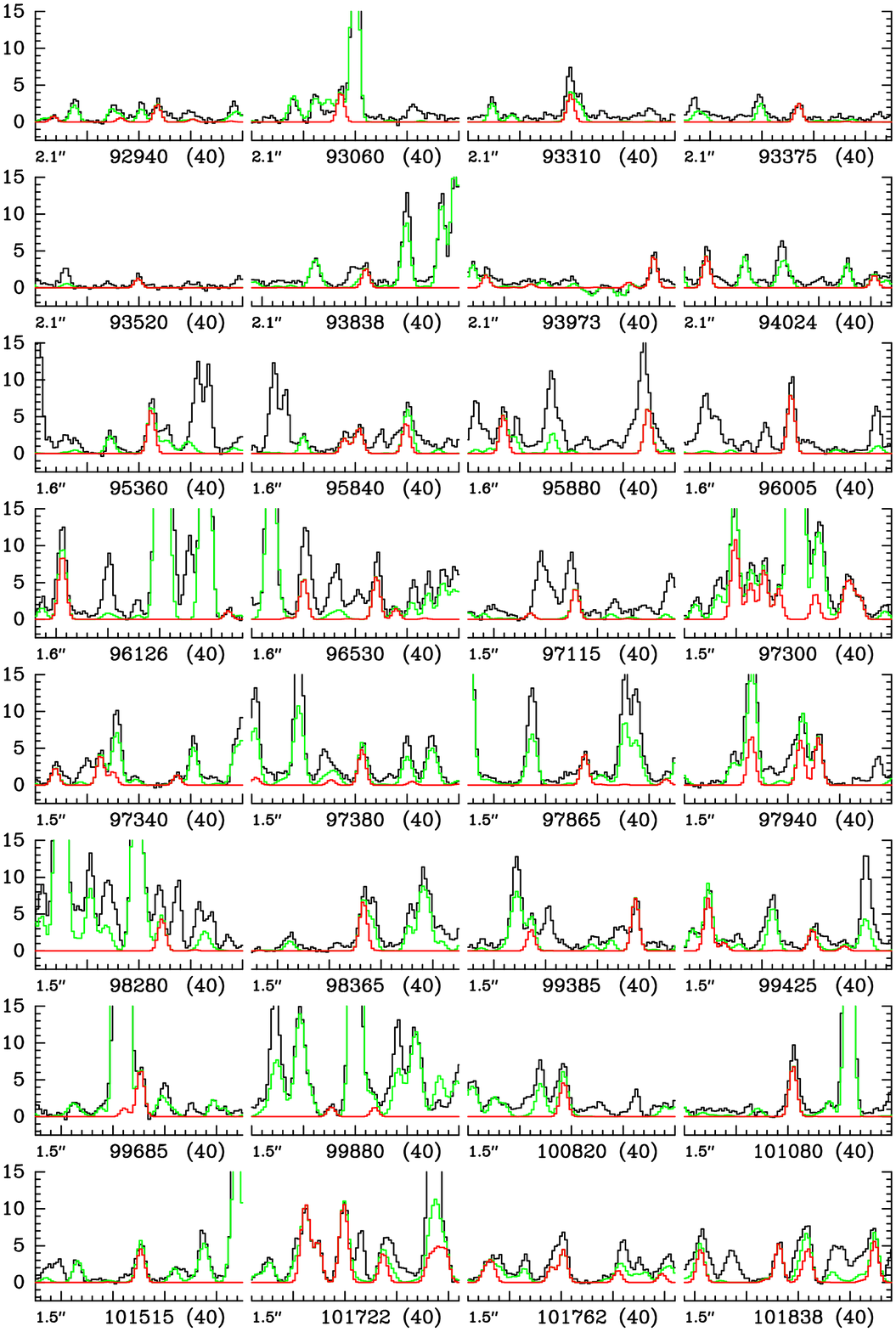}}}
\caption{continued.}
\end{figure}

\begin{figure}
\addtocounter{figure}{-1}
\centerline{\resizebox{0.76\hsize}{!}{\includegraphics[angle=0]{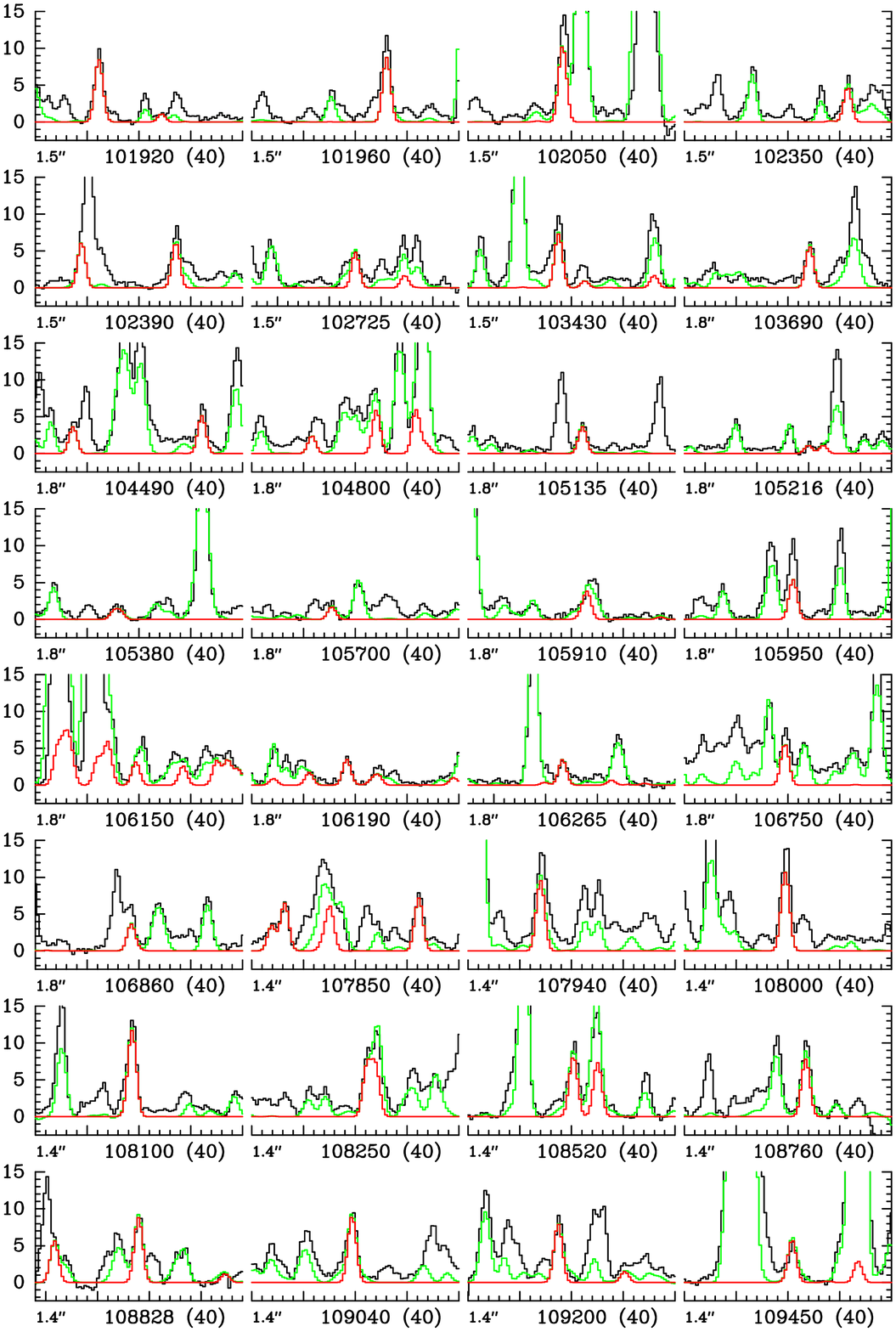}}}
\caption{continued.}
\end{figure}

\begin{figure}
\addtocounter{figure}{-1}
\centerline{\resizebox{0.38\hsize}{!}{\includegraphics[angle=0]{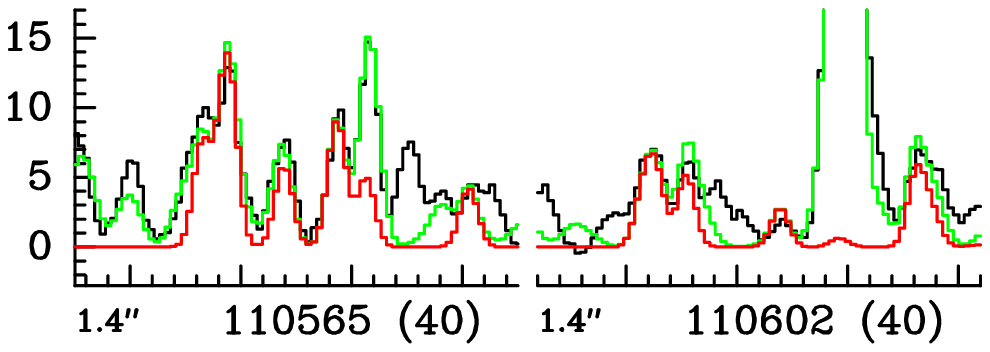}}}
\caption{continued.}
\end{figure}

\clearpage

\begin{figure}
\centerline{\resizebox{0.85\hsize}{!}{\includegraphics[angle=0]{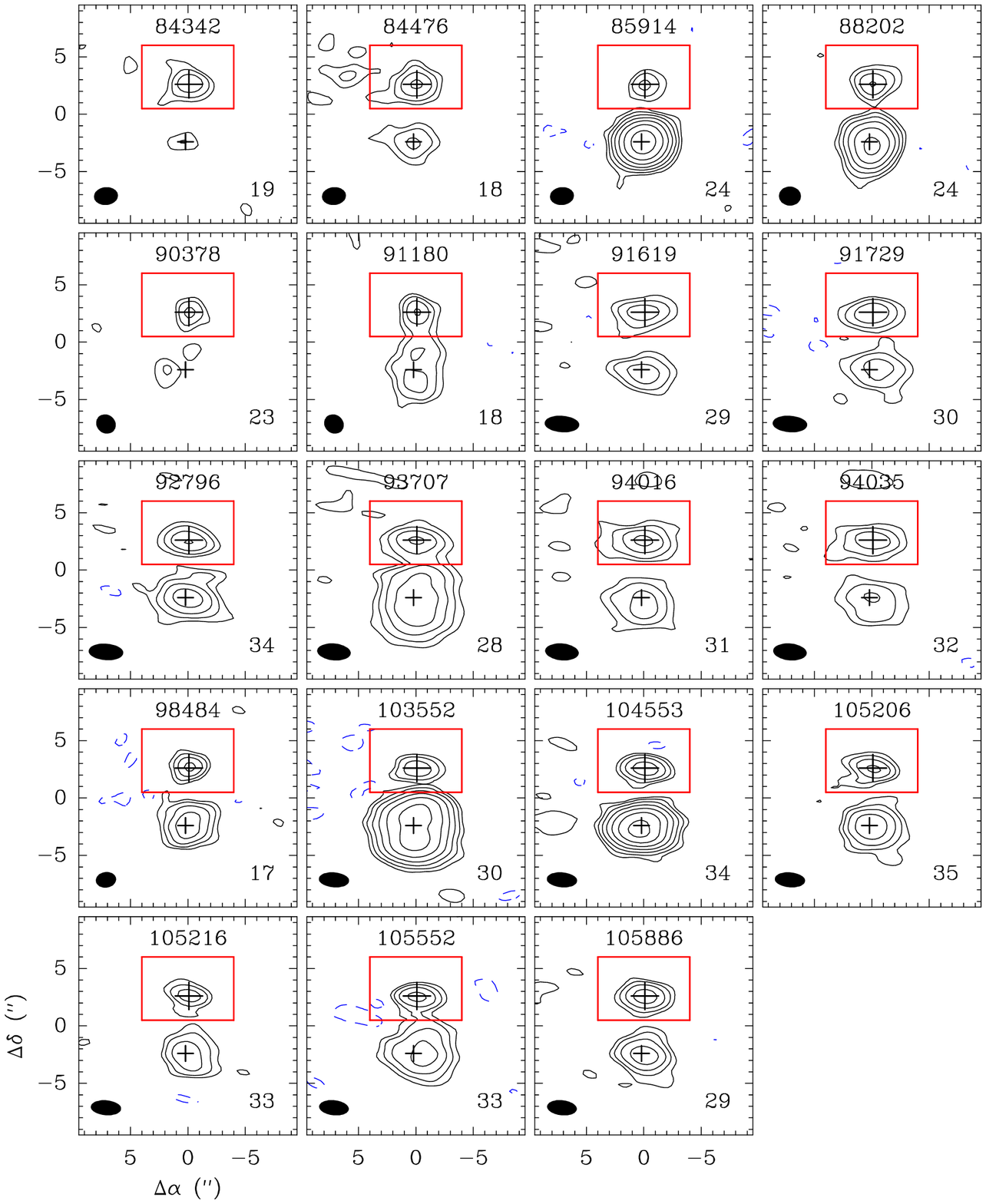}}}
\caption{Integrated intensity maps of selected \textit{i}-PrCN 
transitions toward the northern hot core embedded in Sgr~B2(N). The central
frequency of the integration range in MHz, the synthesized beam (black 
ellipse), and the rms noise 
level $\sigma$ in mJy~beam$^{-1}$~km~s$^{-1}$ are indicated in each panel. The 
negative contour (dashed blue) is $-3\sigma$, and the positive contours (black)
are $2^i\times3\sigma$, with $i$ an integer starting at 0. The big cross 
indicates the position of the northern hot core. The smaller cross
marks the position of the main hot core that has a lower systemic velocity, 
which implies that the 
contours outside the red box do not trace the emission of \textit{i}-PrCN. The 
origin of the offsets corresponds to the phase center 
($\alpha_{\rm J2000}$=17$^{\rm h}$47$^{\rm m}$19.87$^{\rm s}$, 
$\delta_{\rm J2000}$=$-28^\circ$22$'$16$''$).
}
\label{f:maps_i}
\end{figure}

\clearpage

\begin{figure}
\centerline{\resizebox{0.85\hsize}{!}{\includegraphics[angle=0]{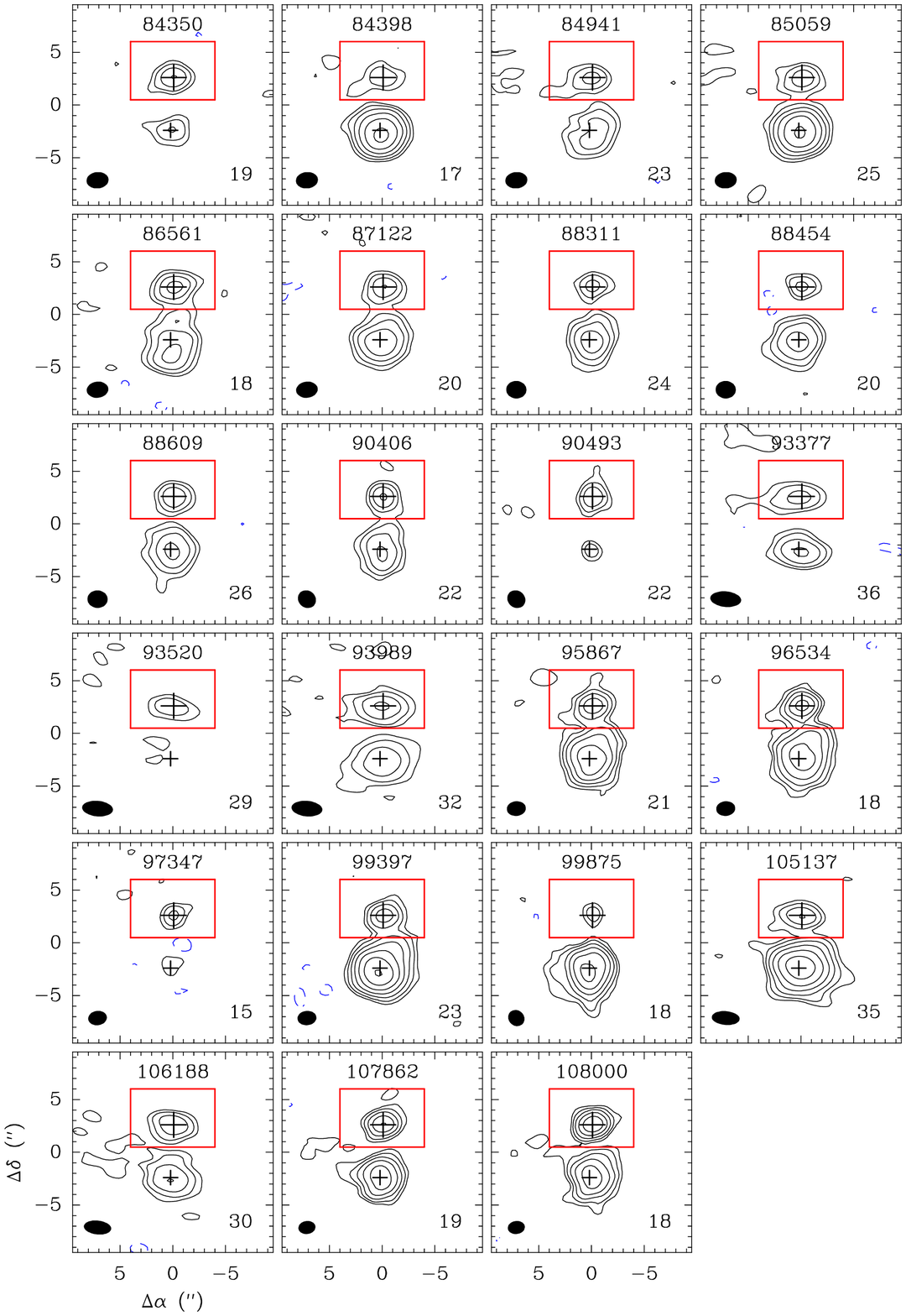}}}
\caption{Integrated intensity maps of selected \textit{n}-PrCN 
transitions toward the northern hot core embedded in Sgr~B2(N). See caption
of Fig.~S3 for more details.}
\label{f:maps_n}
\end{figure}

\clearpage

\begin{figure}
\centerline{\resizebox{0.75\hsize}{!}{\includegraphics[angle=0]{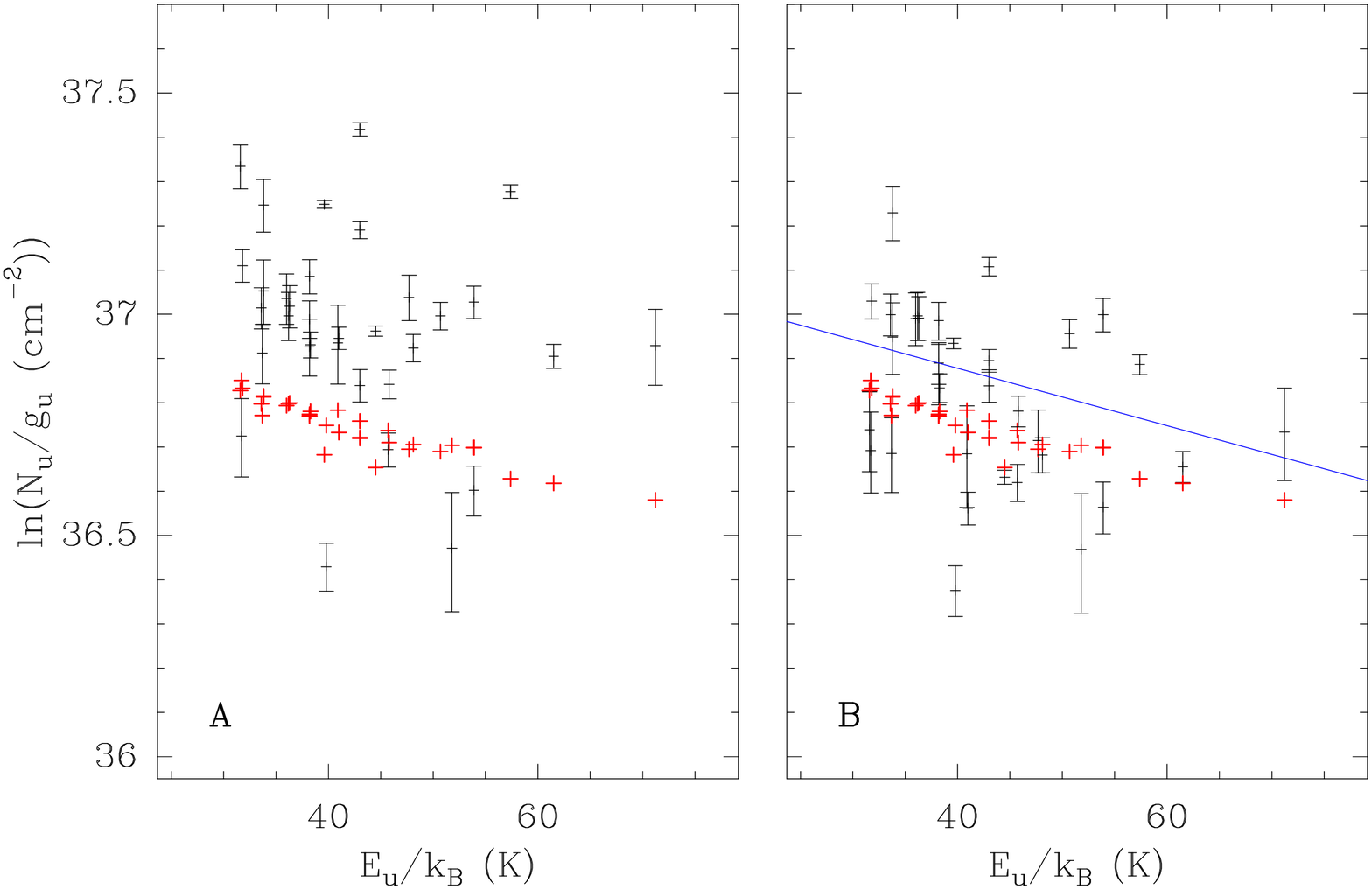}}}
\caption{Population diagram of selected \textit{i}-PrCN transitions 
detected toward the northern hot core embedded in Sgr~B2(N). (\textbf{A}) The 
black crosses with error bars (SD) are derived from the observed 
integrated intensities while the 
red crosses represent the best-fit LTE model of \textit{i}-PrCN. (\textbf{B}) 
Same diagram after removing the contribution of all contaminating molecules 
included in the full LTE model. The blue line is the result of a linear fit to 
the observed datapoints.}
\label{f:popdiag_i}
\end{figure}

\clearpage

\begin{figure}
\centerline{\resizebox{0.75\hsize}{!}{\includegraphics[angle=0]{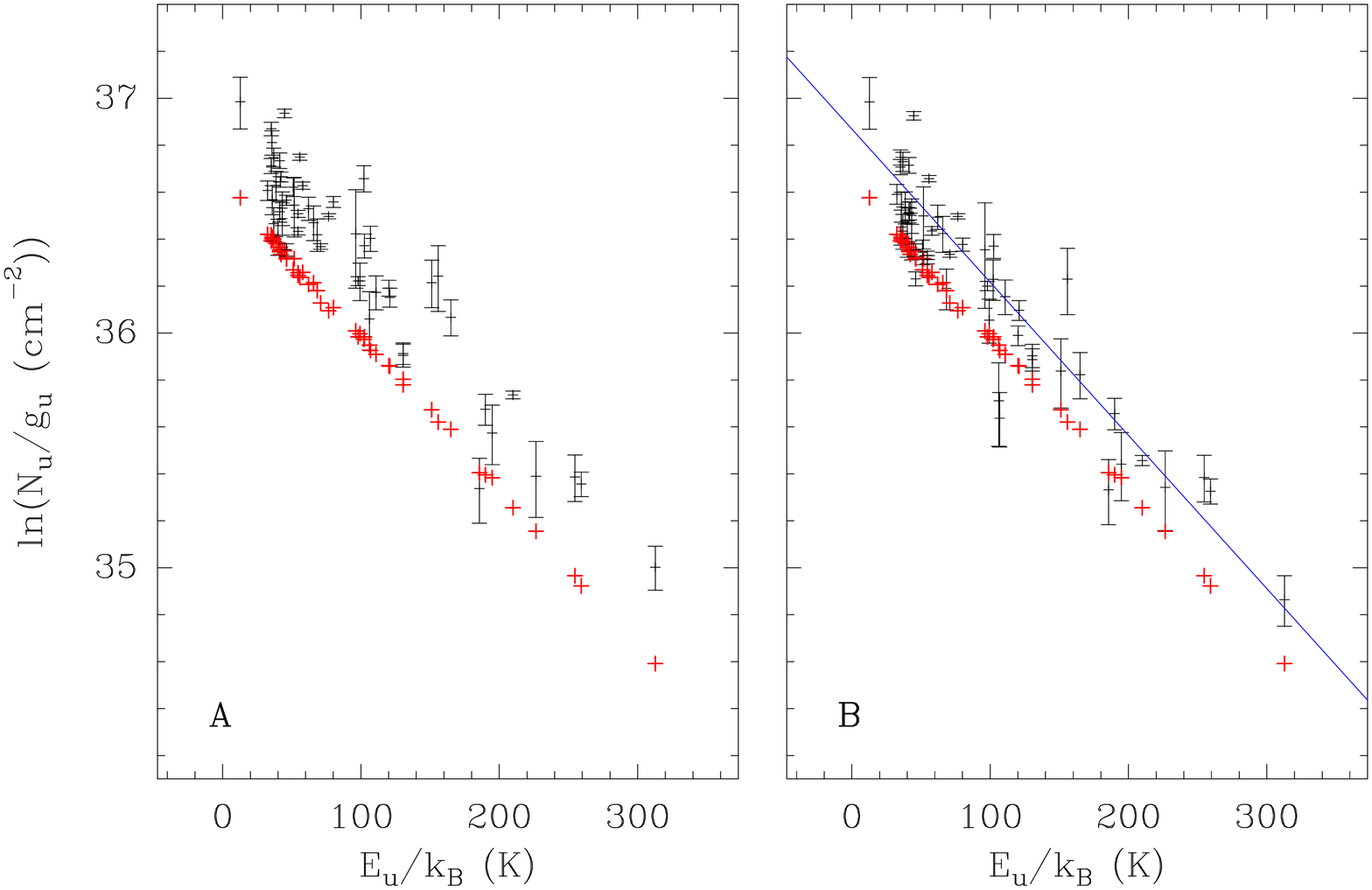}}}
\caption{Population diagram of selected \textit{n}-PrCN transitions 
detected toward the northern hot core embedded in Sgr~B2(N). See caption of
Fig.~S5 for more details.}
\label{f:popdiag_n}
\end{figure}

\clearpage

\begin{table}
\caption{Key surface/mantle reactions in the chemical model leading to 
the production of PrCN.} 
\begin{tabular}{r c l}
\hline
CH$_3$ + CH$_2$CH$_2$CN & $\rightarrow$ & \textit{n}-PrCN\\
C$_2$H$_5$ + CH$_2$CN   & $\rightarrow$ & \textit{n}-PrCN\\
CH$_2$CH$_2$CH$_3$ + CN & $\rightarrow$ & \textit{n}-PrCN\\

CH$_3$ + CH$_3$CHCN     & $\rightarrow$ & \textit{i}-PrCN\\
CH$_3$CHCH$_3$ + CN     & $\rightarrow$ & \textit{i}-PrCN\\
\hline
\end{tabular}
\end{table}

\clearpage

\begin{table}
\caption{Peak fractional abundances of cyanides in the hot-core 
chemical models, and the associated temperatures at which those abundances are 
reached.}
\begin{tabular}{r c c}
\hline
 Molecule, {\em i} & Peak $n(i)/n({\rm H}_{2})$ & Temperature (K) of \\
                   &                           & peak abundance \\
\hline
CH$_3$CN (early) & 2.4e-9 &  94 \\
CH$_3$CN (late)  & 2.6e-9 & 400 \\
C$_2$H$_3$CN     & 2.7e-8 & 301 \\
C$_2$H$_5$CN     & 1.2e-7 & 130 \\
\textit{i}-C$_3$H$_7$CN & 8.0e-9 & 160 \\
\textit{n}-C$_3$H$_7$CN & 3.6e-9 & 160 \\
\hline  
\end{tabular}
\end{table}

\end{document}